\documentstyle[12pt,epsf]{article}

\makeatletter

\@addtoreset{equation}{section}
\makeatother

\newcommand{\drawsquare}[2]{\hbox{%
\rule{#2pt}{#1pt}\hskip-#2pt%
\rule{#1pt}{#2pt}\hskip-#1pt%
\rule[#1pt]{#1pt}{#2pt}}\rule[#1pt]{#2pt}{#2pt}\hskip-#2pt%
\rule{#2pt}{#1pt}}%
\newcommand{\f}{\raisebox{-.5pt}{\drawsquare{6.5}{0.4}}}
\newcommand{\af}{\overline{\f}}

\begin{document}
\begin{titlepage}
\begin{flushright}
{\small KYUSHU-HET-71}\\
hep-ph/0311337
\end{flushright}

\begin{center}
\vspace*{1.5cm}
  
{\Large\bf 
Sparticle Masses in Product-Group Gauge Theories}

\vspace{10mm}
  
{Nobuhito~Maru\footnote{E-mail address: 
maru@postman.riken.go.jp, Special Postdoctoral Researcher} and
Koichi~Yoshioka\footnote{E-mail address:
yoshioka@higgs.phys.kyushu-u.ac.jp}}
\vspace{5mm}

$^*${\it Theoretical Physics Laboratory,\\
RIKEN (The Institute of Physical and Chemical Research),\\
2-1 Hirosawa, Wako, Saitama 351-0198, Japan}\\[2mm]
$^\dagger${\it Department of Physics, Kyushu University, Fukuoka
812-8581, Japan}
\vspace{1cm}

\begin{abstract}\noindent
In this paper, we investigate a possibility to cause supersymmetry
breaking with background modulus fields in product-group gauge
theories. The vacuum expectation values of these moduli are
found to satisfy several relations and the moduli dependences of the
action are fixed by consistency of the model. With this action, we
calculate the mass spectrum of vector and matter multiplets up to
one-loop order of perturbation theory. As an application of the
result, it is found that various properties of higher-dimensional
supersymmetry breaking are well captured in corresponding limits in
the moduli space. In particular, we have finite radiative corrections
to Higgs masses in the case that is closely related to the boundary
condition breaking of supersymmetry.
\end{abstract}
\end{center}
\end{titlepage}
\setcounter{footnote}{0}

\section{Introduction}

Motivated by the gauge hierarchy problem, physics with extra
dimensions has provided new insights on various aspects of particle
physics, cosmology and astrophysics~\cite{ADD,RS}. There however
seems to be some difficulty to discuss the issues in the ultraviolet
regime since higher-dimensional theories are generally
non-renormalizable in the usual sense. Recently, a four-dimensional
framework describing physics of higher-dimensional theories
has been proposed in~\cite{ACG,HPW}. The framework consists of a gauge
theory of product group and scalar fields in the bi-fundamental
representations of the nearest-neighbor gauge groups. If one
assumes that the bi-fundamental scalars develop vacuum expectation
values (VEVs), the gauge groups are broken to a simple diagonal
subgroup. It has been shown that with a sufficiently large number of
gauge groups, the mass spectrum of gauge fields is equivalent in an
intermediate energy regime to Kaluza-Klein (KK) mass spectrum of a
five-dimensional gauge theory. This approach is providing new tools
for four-dimensional model building and moreover for understanding
unexplored properties of higher-dimensional theories. In fact, various
applications along this line has been discussed in the
literature~\cite{apply}.

In Ref.~\cite{KMY}, we have studied a model with supersymmetry (SUSY) 
breaking induced by modulus fields which are naturally incorporated
into the above framework. The modulus fields are found to satisfy some
relations and have non-trivial dependences of the action, if one wants
to obtain a description of higher-dimensional effects. We have
identified the modulus form of the action and verified it for a
five-dimensional vector multiplet by explicitly calculating tree-level
mass spectrum of gaugino and associated adjoint scalar. Various types
of SUSY-breaking mass spectra predicted in higher-dimensional models
appear as corresponding limits in the parameter space of the 
modulus $F$ terms.

In this paper, as an extension of the previous results, we formulate a
four-dimensional model with general matter multiplets and evaluate
one-loop radiative corrections to mass spectrum. We first fix the
modulus couplings to matter multiplets as well as to vector ones, and
then calculate radiative corrections to gaugino and scalar soft
masses in detail. It will be found that the above-mentioned
resemblances to the existing higher-dimensional models still hold even
at quantum level. In addition, we study the ultraviolet behavior of
the radiative corrections and find in some case the corrections to
Higgs masses to be finite. Such a case is explicitly shown to be
closely related to the boundary condition breaking of supersymmetry.

The paper is organized as follows. In Section 2, we first review the
action for vector multiplets and the resulting tree-level mass
spectrum of gauginos and adjoint scalars. We also fix the modulus
couplings to matter multiplets and calculate the (SUSY-breaking)
masses at classical level. With the complete action at hand, radiative
corrections to the masses of vector and matter multiplets are studied
in detail in Section 3. Section 4 gives some arguments about the
finiteness property of loop corrections found in Section 3. There we
will particularly pay attention to some relation between the finite
corrections and the boundary condition breaking of
supersymmetry. Section 5 is devoted to the summary of the paper.

\section{Classical action}

\subsection{Vector multiplets}

The model we consider is a four-dimensional SUSY gauge theory with the
gauge groups $G_1 \times G_2 \times \cdots \times G_N$. We simply
assume all the gauge couplings $g_i$ to have the universal 
value $g$. The four-dimensional $N=1$ vector multiplet $V_i$ of 
the $G_i$ gauge theory contains a gauge field $A_\mu^i$ and a 
gaugino $\lambda^i$. In addition, we have the $N=1$ matter 
multiplets $Q_i$ ($\,i=1,\cdots,N$) in the bi-fundamental
representation, that is, $Q_i$ transforms as $(\f,\,\af)$ under 
the $(G_i,G_{i+1})$ gauge groups. ($G_{N+1}$ is identified 
to $G_1$.) \ The field content of the model is then summarized below:
\begin{eqnarray}
\begin{array}{c|ccccc} 
  & ~G_1~ & ~G_2~ & ~G_3~ & ~\cdots~ & ~G_N~ \\ \hline
  Q_1 & \f & \af & 1 & \cdots & 1 \\
  Q_2 & 1 & \f & \af & \cdots & 1 \\
  \vdots & \vdots & \vdots & \vdots & \ddots & \vdots \\
  Q_N & \af & 1 & 1 & \cdots & \f
\end{array}
\end{eqnarray}
The gauge-invariant Lagrangian for vector multiplets and $Q_i$'s is
written as
\begin{eqnarray}
  {\cal L}_{\rm vec} &=& \sum_i\Bigg[\int d^2\theta\, 
  SW^\alpha_iW_{\alpha i} +\textrm{h.c.}\; 
  +\int d^2\theta d^2\bar\theta\,K_Q(S,S^\dagger)\,
  Q_i^\dagger e^{\sum V} Q_i\Bigg], 
  \label{Lvec}
\end{eqnarray}
where $W^\alpha_i$ is the field strength superfield of the $G_i$ gauge
group, and $S$ is a dilaton-like background superfield whose scalar
component determines a value of $g$ [see Eq.~(\ref{S})]. In fact, one
may introduce a dilaton field for each gauge theory $G_i$. We now take
the universal value of gauge couplings and use $S$ as a representative
of such dilaton fields. The normalization function $K_Q(S,S^\dag)$ of
the matter field $Q_i$ is found~\cite{KMY}
\begin{equation}
  K_Q(S,S^\dag) \;=\; \frac{8}{1/S+1/S^\dag},
  \label{K}
\end{equation}
which can be fixed by several classical-level consistencies as will be
discussed at the end of this section. Moreover, this form of $K_Q$ will
be found in the next section to be also important for loop-level
consistency of the model.

If one considers the case that $Q_i$'s have VEVs proportional to the
unit matrix, $\langle Q_i\rangle=v$,\footnote{More generic cases with
non-universal values of the gauge couplings and VEVs of $Q$'s
correspond to gauge theories on curved backgrounds~\cite{warp} and
low-energy gauge symmetry breaking~\cite{inst}. We do not consider
these possibilities in the present work.} the gauge symmetries are
broken down to its diagonal subgroup $G$. With this symmetry breaking,
the mass eigenvalues and eigenstates of the gauge fields are given by
\begin{equation}
\begin{array}{l}
  \displaystyle{m_n \;=\; 2v\sin\frac{n\pi}{N},} \\[2mm]
  \displaystyle{\widetilde A_\mu^n \;=\; \frac{1}{\sqrt{N}} 
  \sum_{j=1}^N (\omega_n)^j A_\mu^j,}
\end{array} \qquad (n=0,\cdots,N-1)
\label{KKmass} 
\end{equation}
where $\omega_n=e^{2\pi in/N}$. We thus find the KK tower of massive
gauge fields as well as a massless one which belongs to the unbroken
subgroup $G$. It is interesting to note that the spectrum matches to
that of a five-dimensional gauge theory with a compact extra
dimensions~\cite{ACG,HPW}. That is, in the large $N$ limit with $v/N$
fixed, the gauge boson masses approximately become
\begin{equation}
  m_n \;\simeq\; \frac{2vn\pi}{N} \;=\; \frac{n}{R},
  \label{KKmasap}
\end{equation}
where the compactification radius $R$ is identified 
as $2\pi R=N/v$. With this identification, $v$ is interpreted as an
ultraviolet cutoff of the effective four-dimensional theory, and $N$
is the number of KK modes which exist between the cutoff and
compactification scales.

In the absence of SUSY breaking, the gauginos and the associated
adjoint scalars, which are linear combinations of the original 
fields $\lambda_i$ and $Q_i$, have degenerate mass spectrum with that
of the gauge bosons (\ref{KKmass}). Before examining SUSY-breaking
mass splitting, it may be instructive to recall the modulus fields in
our model and their mutual relations. In addition to $S$ introduced
above, there is a modulus $Q$ which determines the size of the 
universal VEV $v$. The modulus $Q$ may be a normalized composite
field or a linear combination of $Q_i$'s. Throughout this paper, we
assume just for simplicity that the moduli forms of $S_i$ and $Q_i$
are invariant under the ``translation'' transverse to the four
dimensions, that is, they are $i$-independent. The forms of VEVs of
these moduli are then defined as
\begin{eqnarray}
  S \;=\; \frac{1}{4g^2} + F_S \theta^2, \label{S}\qquad
  Q \;=\; v + F_Q \theta^2.
\end{eqnarray}
Furthermore, it will turn out to be useful to define the additional
background modulus fields with the following VEVs
\begin{eqnarray}
  S_4 \;=\; \frac{1}{4g_4^2} + F_{S4} \theta^2, \qquad
  S_5 \;=\; \frac{1}{4g_5^2} + F_{S5} \theta^2, \qquad
  T \;=\; \frac{1}{R} + F_T \theta^2,
\end{eqnarray}
where $g_4$ and $g_5$ are the effective gauge coupling constants in
four and five dimensions, respectively. It is important to notice that
all these modulus fields are not independent variables. By comparing
the Kaluza-Klein theory with the low-energy description of the model
below the cutoff $v$, one finds the following matching conditions 
between the parameters~\cite{ACG,HPW}:
\begin{eqnarray}
  2\pi R \;=\; \frac{N}{v}, \qquad
  g_4^2 \;=\; \frac{g^2}{N}.
  \label{match}
\end{eqnarray}
The first condition in (\ref{match}) was already adopted in
(\ref{KKmasap}), which was required to match the spectrum to that of
KK theory. The second condition is regarded as a volume suppression of
bulk gauge coupling in compactifying an extra dimension. In addition
to these, the normalization of gauge kinetic terms provides a relation
between $g_4$ and $g_5$, irrespectively of how to define a
five-dimensional model,
\begin{eqnarray}
  g_5^2 &=& 2\pi Rg_4^2.
  \label{5T4}
\end{eqnarray}
These three relations among the couplings suggest the following
relations among the modulus fields:\footnote{The 1PI and holomorphic
gauge couplings differ only at higher-loop level in perturbation theory.}
\begin{eqnarray}
  S_4 \;=\; N S, \qquad
  S_5 \;=\; QS, \qquad
  T \;=\; \frac{2\pi}{N}Q.
  \label{TQ}
\end{eqnarray}
We thus find that the modulus fields $S_4$, $S_5$ and $T$ are 
expressed in terms of two moduli $S$ and $Q$. As a result, the
two-dimensional parameter space of the $F$-components of $S$ and $Q$
describes SUSY-breaking patterns in our model. In Ref.~\cite{KMY}, we
clarified that several limits in this parameter space describe bulk
SUSY-breaking patterns which have been discussed in the
literature.\footnote{Supersymmetry breaking that is local in the
higher-dimensional bulk was studied within four-dimensional
product-group gauge theories~\cite{gaudecon}.} In Table~\ref{Fterm},
the typical cases are presented for the dilaton/moduli dominated
SUSY-breaking~\cite{sugra} and for SUSY breaking by the radius 
modulus $F$ term~\cite{noscale,radion}. The specification of each
limit is also given in the table. For example, the dilaton dominance
scenario is defined by $F_T=0$, which is in turn translated to the
limit $F_Q=0$ in our model [see Eq.~(\ref{TQ})].
\begin{table}[htbp]
{\small
\begin{center}
\begin{tabular}{lcccccc} \hline\hline
  && $F_S$ & ~$F_Q$~ & $F_{S4}$ & $F_{S5}$ & $F_T$ \\ \hline
  \parbox{18mm}{{\normalsize dilaton}\\ ($F_T\equiv 0$)} &&
  $F_S$ & $0$ & $NF_S$ & $vF_S$ & 0 \\[3mm]
  \parbox{18mm}{{\normalsize moduli}\\ ($F_{S4}\equiv 0$)} && 
  0 & $F_Q$ & 0 & $\displaystyle{\frac{1}{4g^2}F_Q}$ & 
  $\displaystyle{\frac{2\pi}{N}F_Q}$ \\[3mm]
  \parbox{18mm}{{\normalsize radion}\\ ($F_{S5}\equiv 0$)} &&
  $\displaystyle{\frac{-F_Q}{4g^2v}}$ & $F_Q$ & 
  $\displaystyle{\frac{-NF_Q}{4g^2v}}$ & 0 & 
  $\displaystyle{\frac{2\pi}{N}F_Q}$ \\[3mm] \hline\hline
\end{tabular}
\caption{{\normalsize The modulus $F$-terms of the typical
SUSY-breaking scenarios.}}
\label{Fterm}
\end{center}}
\end{table}
In the following, we will study the SUSY-breaking effects from these
modulus fields and examine sparticle spectrum of the model.

If one introduces suitable potential terms for the modulus fields,
their VEVs are fixed to some point in the parameter space. For
example, since $S$ is a dilaton for each gauge group, stabilization
mechanisms proposed in the literature could be incorporated in the
present model. The situation is similar for the modulus $Q$. Moreover
in properly describing five-dimensional theory, $Q$ is assumed to be
stabilized by relevant potential terms~\cite{ACG,gaudecon}. However
details of potential form are not relevant to the present
analysis. Without referring to specific models, we explore the whole
parameter space of the modulus $F$-terms and then examine several
limits corresponding to bulk SUSY-breaking patterns. Moreover we do
not try to construct specific dynamics for the modulus fields where
potential couplings are tuned for realizing the fifth dimension. Our
aim here is not to present five-dimensional theories. It is only the
relevant region of moduli space where our model reproduces bulk
SUSY-breaking scenarios. In other words, the present framework
contains unexplored four-dimensional phenomena of SUSY breaking.

\medskip

Now the SUSY-breaking mass spectrum can be derived from the Lagrangian
(\ref{Lvec}) with turning on the modulus $F$ terms. The result is 
written by use of the above-defined moduli and their relations;
\begin{eqnarray}
  m_{\lambda_{n\pm}} &=& \pm\sqrt{m_n^2+\left|
  \frac{F_{S5}}{2\langle S_5\rangle}\right|^2\,}
  -\frac{F_T}{2\langle T\rangle}, 
  \label{mlam} \\
  m_{c_n}^2 &=& m_n^2 +2\,\textrm{Re}
  \left(\frac{F_{S5}^{\,*}}{\langle S_5\rangle}
  \frac{F_T}{\langle T\rangle}\right),
  \label{cn}
\end{eqnarray}
where $m_{\lambda_{n\pm}}$ and $m_{c_n}$ are the KK masses of gauge
fermions and adjoint scalars. The bracket $\langle~\rangle$ denotes a
VEV of the lowest scalar component. The positive (negative) sign 
in $m_{\lambda_{n\pm}}$ corresponds to the gaugino (the goldstone
fermion) masses in the supersymmetric limit. Note that the results are
expressed by five-dimensional quantities only. In particular, it is
found the zero-mode gaugino mass is given by {\it both} the radius
modulus and the five-dimensional dilaton. This result is expected from
the equation (\ref{5T4}) which implies $S_4$ depends both on $T$ 
and $S_5$ \ ($TS_4=2\pi S_5$).

We here explicitly show the several limits in order. As mentioned
before, the dilaton-dominated SUSY breaking is characterized by the
condition $F_T=0$. In this limit, we find 
\begin{equation}
  m_{\lambda_{n\pm}} \;=\; \pm\sqrt{m_n^2+|2g^2F_S|^2 }, \qquad
  m_{c_n}^2 \;=\; m_n^2. \qquad\quad [{\rm dilaton}]
  \label{dg}
\end{equation}
The gaugino mass spectrum is indeed the one expected in supergravity
models. All the KK states including the zero mode receive the
universal SUSY-breaking contribution from the modulus $S$, the two
level--$n$ spinors are degenerate in mass, and the mass splitting
between bosons and fermions are equal for all KK modes. These results
may be understood from the fact that the dilaton field commonly
couples to any field in the theory.

On the other hand, the limit of the moduli-dominated SUSY breaking is
defined by $F_{S_4}=0$ and then leads to
\begin{equation}
  m_{\lambda_{n\pm}} \;=\; \pm\sqrt{m_n^2+\left|\frac{F_Q}{2v}\right|^2}
  -\frac{F_Q}{2v}, \qquad
  m_{c_n}^2 \;=\; m_n^2 +2\left|\frac{F_Q}{v}\right|^2.
  \qquad\quad [{\rm moduli}]
\end{equation}
It is interesting to note that even when the SUSY-breaking effect is
turned on, the zero-mode gaugino remains massless.\footnote{The level
$n=0$ spinor being affected by the non-zero $F$-terms is the goldstone
fermion associated to gauge symmetry breaking.}
This is exactly the tree-level spectrum predicted in this class of
SUSY-breaking models~\cite{sugra,KK}. Since $F_{S4}=0$ by definition,
the zero-mode gaugino mass is vanishing and is shifted at loop level
by string threshold corrections or effects of bulk fields. The
situation is also similar to the model where vector multiplets
behave as messengers of SUSY breaking and sparticle soft masses are
calculated from the wave function renormalization in four
dimensions~\cite{GR}. There may be an intuitive explanation for
the above type of spectrum. That is, a nonzero $F$ term of the modulus
which determines KK masses does not induce tree-level SUSY-breaking
masses for zero modes. This is because these two mass terms are
proportional to KK numbers. In our case, such a modulus corresponds to
the one whose scalar component obtains a VEV $\propto 1/R$ and is
given by $Q\propto T$.

The last example $F_{S_5}=0$ realizes the SUSY breaking with the
radius modulus field. The sparticle mass spectrum in this limit becomes 
\begin{equation}
  m_{\lambda_{n\pm}} \;=\; \pm m_n -\frac{F_T}{2\langle T \rangle}, 
  \qquad m^2_{c_n} \;=\; m_n^2. \qquad\quad [{\rm radion}]
  \label{gaurad}
\end{equation}
The gaugino mass matches to that calculated in~\cite{radion}. The
vanishing SUSY-breaking masses of the adjoint scalars agree with the
fact that this limit is equivalent to the Scherk-Schwarz
mechanism~\cite{SS}, which is now applied to the $SU(2)_R$ symmetry
under which the adjoint scalar is singlet and does not get a
symmetry breaking mass.

\medskip

Before closing this subsection, we comment on the normalization
function $K_Q(S,S^\dagger)$. The form of $K_Q$ (\ref{K}) is determined
so that it satisfies several nontrivial requirements. First, the
holomorphy requires the normalization of $Q_i$ to 
be $\langle K_Q\rangle=1/g^2$, which leads to the same normalization
for the vector and adjoint chiral multiplets of the low-energy $G$
gauge theory. Moreover, the radius superfield in our model becomes
independent of the dilaton superfield [see (\ref{TQ})]. It seems 
plausible since an undesirable relation between the theta angle and
the graviphoton field does not arise. Another consistency is 
about the 5-5 component of five-dimensional metric $g_{55}$. In a
continuum five-dimensional theory, the kinetic terms of bosonic fields
along the fifth dimension have a dependence of $g_{55}$ 
as $\sqrt{g_{55}}\,g^{55}\,\propto\,1/R$. In the present model, the
second term in the Lagrangian (\ref{Lvec}) becomes the kinetic energy
in the continuum limit and its modulus dependence is given by 
$\langle K_Q(S,S^\dagger)\,Q^\dagger Q\rangle$. The 
equation (\ref{K}) then 
indicates $\langle K_QQ^2\rangle$ $\sim$ $\langle SQ^2\rangle$
$\sim$ $\langle S_5T\rangle$. Consequently, the metric dependence
agrees for a fixed value of the five-dimensional coupling $g_5$. 

\subsection{Matter multiplets}

Next let us discuss matter multiplets, which are regarded as
hypermultiplets if one takes the five-dimensional limit. As in the
case of vector multiplets, we will determine the proper form of moduli
dependences of the matter action and then examine SUSY-breaking mass
spectrum at tree level. One-loop corrections to the masses of scalar
components in these matter multiplets will be studied in Section 3.

In addition to the fields in the previous subsection, we introduce a
set of vector-like chiral multiplets $\Phi_i$ and $\bar\Phi_i$ for
each gauge group $G_i$:
\begin{eqnarray}
 \begin{array}{c|ccccc}
   & ~G_1~ & ~G_2~ & ~G_3~ & ~\cdots~ & ~G_N~ \\ \hline
  \Phi_1 & \f & 1 & \cdots & \cdots & 1 \\
  \bar\Phi_1 & \af & 1 & \cdots & \cdots & 1 \\
  \Phi_2 & 1 & \f & \cdots & \cdots & 1 \\
  \bar\Phi_2 & 1 & \af & \cdots & \cdots & 1 \\
  \vdots & \vdots & \vdots & \vdots & \ddots & \vdots \\
  \Phi_N & 1 & 1 & \cdots & \cdots & \f \\
  \bar\Phi_N & 1 & 1 & \cdots & \cdots & \af \\
 \end{array}
\end{eqnarray}
The gauge-invariant Lagrangian for the matter sector can be described by
\begin{eqnarray}
  {\cal L}_{\rm mat} &=& \sum_{i=1}^N\Bigg[\int d^2\theta d^2\bar\theta\,
  K_h(S,S^\dagger) \Big[\Phi_i^\dagger e^{2V_i} \Phi_i 
  +\bar\Phi_i e^{-2V_i} \bar\Phi_i^\dagger\Big] \nonumber \\
  && \hspace{2cm} +\int d^2\theta\, \Big[Y\bar\Phi_i Q_i \Phi_{i+1} 
  +Z\bar\Phi_i\Phi_i\Big] +{\rm h.c.} \Bigg],
 \label{Lmat}
\end{eqnarray}
where $\Phi_{N+1}$ ($\bar\Phi_{N+1}$) is identified 
to $\Phi_1$ ($\bar\Phi_1$). We have assumed the coupling constants are
universal, for a simplicity. The background chiral superfields $Y$ 
and $Z$ in the superpotential are the modulus fields providing Yukawa
and mass parameters. As will be seen below, these fields are expressed
in terms of the moduli $S$ and $Q$. Similar to the vector multiplet
case, it is a non-trivial problem to fix the moduli dependence of the
normalization function $K_h(S,S^\dag)$. That issue will be discussed
later.

\subsubsection{Superpotential}

First we study the moduli dependence of the superpotential terms. It
is convenient to rescale the matter multiplets 
as $(\Phi,\bar\Phi)\to\langle K_h\rangle^{-\frac{1}{2}}(\Phi,\bar\Phi)$
so that the kinetic terms are canonical. In the rescaled basis, the
superpotential contribution to supersymmetric mass terms reads
\begin{eqnarray}
  W_{\rm mass} &=& \frac{1}{\langle K_h\rangle} \sum_{i,j}\bar\Phi_i
  \Big(\langle Y\rangle v\,\delta_{i+1,j} 
  +\langle Z\rangle\delta_{i,j}\Big)\Phi_j.
\end{eqnarray}
It is easily found that the 
relation $\langle Y\rangle=\sqrt{2}\langle K_h\rangle$ must be
satisfied if one requires the matter spectrum to be equivalent to that
of vector multiplet (i.e.\ that of KK theory in the five-dimensional
limit). This implies the moduli form 
\begin{equation}
  Y \;=\; K_h(S,S).
  \label{Y}
\end{equation}
Given this relation, diagonalizing the supersymmetric mass term leads to
\begin{eqnarray}
  W_{\rm mass} &=& m_n\widetilde{\bar\Phi}_m \delta_{mn}
  \widetilde\Phi_n +\big(\langle Z\rangle/\langle K_h\rangle +v\big)\,
  \widetilde{\bar\Phi}_n\widetilde\Phi_n, 
  \label{mat-mass}
\end{eqnarray}
where the supersymmetric eigenstates $\widetilde\Phi_n$ 
and $\widetilde{\bar\Phi}_n$ ($m,n=0,\cdots,N-1$) are defined similar
to those of vector fields (\ref{KKmass}). The irrelevant phase
factors have been absorbed with field redefinitions so that the mass 
eigenvalues are real. In the absence of nonzero $F$-terms, the mass
eigenstates take the same form as the gauge fields. The first term in
the right-handed side of (\ref{mat-mass}) corresponds to the KK masses
and the second one to a bare mass parameter of matter multiplet. When 
the bare mass $m$ is defined 
as $\langle Z\rangle/\langle K_h\rangle+v=m/R$, the modulus $Z$ should
take the form
\begin{equation}
  Z \;=\; \bigg(\frac{2\pi m}{N}-1\bigg)Q K_h(S,S). 
  \label{Z}
\end{equation}
Now the moduli dependences of the superpotential part are fully
determined, their contribution to SUSY-breaking masses can be
analyzed. From (\ref{Lmat}), (\ref{Y}) and (\ref{Z}), we find the
mass eigenvalues of spinor components
\begin{eqnarray}
  m_{\psi_n} &=& m_n + \frac{m}{R}, \qquad (n=0,\cdots,N-1).
  \label{spimass}
\end{eqnarray}
The scalar soft mass terms from the superpotential are written as
\begin{equation}
  {\cal L}_{\rm W} \;=\; \left[\bigg\langle
  \frac{\partial\ln K_h(S,S)}{\partial \ln S}\bigg\rangle
  \frac{F_S}{\langle S\rangle}+\frac{F_T}{\langle T\rangle}\right]
  \widetilde{\bar\phi}_m \bigg(m_n+\frac{m}{R}\bigg)\delta_{mn}
  \widetilde\phi_n + {\rm h.c.},
  \label{sp}
\end{equation}
where $\phi$ and $\bar\phi$ are the scalar components of $\Phi$ 
and $\bar\Phi$, respectively. This is the holomorphic mixing mass term
of $\phi$ and $\bar\phi$. In the rescaled basis, the mixing masses
depend on the K\"ahler factor $K_h$ but they should be cancelled away
if one includes full contribution of modulus fields as will be seen
below.

\subsubsection{K\"ahler potential}

Scalar masses also come from the K\"ahler potential since it has
the moduli dependence and so non-canonical. The masses of the spinor
components (\ref{spimass}) does not get changed by the K\"ahler
terms. For the rescaled fields defined above, we read off the
Lagrangian (\ref{Lmat}),
\begin{eqnarray}
  {\cal L}_{\rm K} &\!\!=& \sum_{i=1}^N \Bigg[ |F_{\phi_i}|^2
  +\bigg\langle\frac{\partial \ln K_h}{\partial S}\bigg\rangle F_S 
  F_{\phi_i}^\dagger\phi_i
  +\bigg\langle\frac{\partial \ln K_h}{\partial S^\dagger}\bigg\rangle
  F_S^\dagger \phi_i^\dagger F_{\phi_i} 
  +\bigg\langle\frac{1}{K_h}
  \frac{\partial^2 K_h}{\partial S\partial S^\dagger}\bigg\rangle
  |F_S|^2 \phi_i^\dagger \phi_i \nonumber \\[1mm] 
  && +|F_{\bar\phi_i}|^2 +\bigg\langle\frac{\partial \ln K_h}{\partial S}
  \bigg\rangle F_S\bar\phi_i F_{\bar\phi_i}^\dagger
  +\bigg\langle\frac{\partial \ln K_h}{\partial S^\dagger}\bigg\rangle
  F_S^\dagger F_{\bar\phi_i} \bar\phi_i^\dagger
  +\bigg\langle\frac{1}{K_h}
  \frac{\partial^2 K_h}{\partial S \partial S^\dagger}\bigg\rangle
  |F_S|^2 \bar\phi_i \bar\phi_i^\dagger \Bigg]. \nonumber \\
\end{eqnarray}
Integrating out the matter auxiliary components $F_{\phi_i}$ 
and $F_{\bar\phi_i}$ and moving to the supersymmetric mass basis, we have 
\begin{eqnarray}
  {\cal L}_{\rm K} &=& -\sum_{n=0}^{N-1} \Bigg[\bigg(|m_{\psi_n}|^2+
  \bigg\langle\frac{\partial \ln K_h}{\partial S}
  \frac{\partial \ln K_h}{\partial S^\dagger} -\frac{1}{K_h}
  \frac{\partial^2 K_h}{\partial S \partial S^\dagger}\bigg\rangle 
  F_SF_S^\dagger \bigg) \Big(\widetilde\phi_n^\dagger\widetilde\phi_n
  +\widetilde{\bar\phi}_n\widetilde{\bar\phi}_n^\dagger \Big) 
  \nonumber \\[2mm] 
  && \hspace{5cm} +2m_{\psi_n}\bigg\langle
  \frac{\partial \ln K_h}{\partial S}\bigg\rangle F_S\, 
  \widetilde{\bar\phi}_n\widetilde\phi_n +{\rm h.c.} \Bigg].
\end{eqnarray}
Combining this with the superpotential contribution to scalar masses
(\ref{sp}) and noting that
$\langle\partial\ln K_h(S,S)/\partial S\rangle =2\langle
\partial\ln K_h/\partial S\rangle=2\langle
\partial\ln K_h/\partial S^\dagger\rangle$, we finally obtain the
scalar mass matrix for matter multiplets
\begin{eqnarray}
  {\cal L}_{\rm scalar} &=& -\sum_{n=0}^{N-1}
  \pmatrix{\widetilde{\phi}_n^\dagger & \widetilde{\bar\phi}_n \cr}
  \pmatrix{ |m_{\psi_n}|^2 +\Gamma\left|
    \displaystyle{\frac{F_S}{\langle S\rangle}}\right|^2 & -m^*_{\psi_n} 
    \displaystyle{\frac{F_T^{\,*}}{\langle T\rangle}} \cr 
    -m_{\psi_n}\displaystyle{\frac{F_T}{\langle T\rangle}} & 
    |m_{\psi_n}|^2 +\Gamma\left|
    \displaystyle{\frac{F_S}{\langle S\rangle}}\right|^2 }
  \pmatrix{\widetilde{\phi}_n \cr \widetilde{\bar\phi}_n\!{}^\dagger \cr},
  \\[3mm]
  && \hspace{2.7cm} \Gamma \;\equiv\; \bigg\langle 
  \frac{\partial^2 \ln K_h^{-1}}{\partial \ln S\partial \ln S^\dagger}
  \bigg\rangle.
\end{eqnarray}
Thus the tree-level mass eigenvalues are given by
\begin{eqnarray}
  m^2_{\phi_{n\pm}} &=& |m_{\psi_n}|^2
  \pm m_{\psi_n}\frac{F_T}{\langle T\rangle}
  +\Gamma\left|\frac{F_S}{\langle S\rangle}\right|^2, \qquad 
  (n=0,\cdots,N-1).
  \label{stree}
\end{eqnarray}
One can see in this formula that the effect of the modulus $S$ is
controlled by the factor $\Gamma$, which is a function of the
normalization constant in the matter K\"ahler term, while the $F_T$
part does not depend on the K\"ahler normalization. The 
factor $\Gamma$ becomes, for example, $\Gamma=b/4$ for the K\"ahler form 
\begin{equation}
  K_h \;=\; (S+S^\dagger)^b X(S)X(S^\dagger),
  \label{kh}
\end{equation}
where $b$ is a constant and $X$ is an arbitrary function. We will find
that $b=1$ and $X=$ constant are the appropriate form if one wanted to 
describe five-dimensional theory. This choice is supported by examining
tree-level spectrum and radiative corrections, which will be discussed
in the following sections. Such K\"ahler function indeed satisfies the
following non-trivial consistencies that (i) the mass spectrum of
matter scalars including SUSY-breaking effects coincides with that of
gauginos, (ii) the cutoff dependences of radiative corrections become
consistent with known results, and (iii) the moduli dependence of the
action has the proper form similar to the argument given at the end of
Section~2.1.

\subsubsection{Various limits}

We have determined all the moduli dependences of the action except 
for $K_h$ and presented the superparticle spectrum for vector and
matter multiplets in four-dimensions: (\ref{KKmass}), (\ref{mlam}),
(\ref{cn}), (\ref{spimass}), and (\ref{stree}). It has been shown 
for the gaugino mass that several limits of the $F$ terms suggest the
five-dimensional properties of SUSY breaking. Let us examine 
scalar mass spectrum for the typical cases presented in
Table~\ref{Fterm}. The first is the dilaton dominance limit defined 
by $F_T=0$. It is found from (\ref{stree}) the scalar mass eigenvalues
in this limit become
\begin{equation}
  m^2_{\phi_{n\pm}} \;=\; |m_{\psi_n}|^2 
  +\Gamma\left|\frac{F_S}{\langle S\rangle }\right|^2. \qquad\quad 
  [{\rm dilaton}]
  \label{ds}
\end{equation}
As expected, the SUSY-breaking contributions are universal for all 
scalar fields. Given the K\"ahler form (\ref{kh}) with $b=1$, the
spectrum (\ref{ds}) (with a vanishing bare mass $m$) is identical to
that of the gauge fermions (\ref{dg}). The second limit we consider is
the moduli dominance which leads to the mass eigenvalues
\begin{equation}
  m^2_{\phi_{n\pm}} \;=\; |m_{\psi_n}|^2 \pm m_{\psi_n}\frac{F_Q}{v}. 
  \qquad\quad [{\rm moduli}]
\end{equation}
It is interesting to note that the spectrum is predicted independently
of detailed form of the matter K\"ahler function $K_h$. Furthermore
the zero mode ($n=0$) does not get SUSY-breaking effects and remains
massless. These features are certainly shared with gauginos. In the
approximation that SUSY-breaking effect is much smaller than the
compactification scale, the mass eigenvalues of the excited modes is
written by $m_{\phi_{n\pm}}\simeq m_{\psi_n} \pm F_Q/2v$, which is
also consistent to that of KK gauginos.

In the limit $F_{S5}=0$, we have
\begin{equation}
  m^2_{\phi_{n\pm}} \;=\; |m_{\psi_n}|^2 
  \pm m_{\psi_n}\frac{F_T}{\langle T\rangle}
  +\Gamma\left|\frac{F_T}{\langle T\rangle}\right|^2. \qquad\quad
  [{\rm radion}]
  \label{rs}
\end{equation}
The mass eigenvalues (\ref{rs}) have the same form as those of
gauginos when $b=1$ for the matter K\"ahler form. Note that the 
masses of the excited modes approximately agree with the moduli
dominance limit. The only difference is whether the zero mode is
massless or not, which mode is identified to the low-energy degree of
freedom in the five-dimensional viewpoint.

We thus find from the tree-level analysis of mass spectrum that $b=1$
in the K\"ahler form (\ref{kh}) is a suitable choice for the
normalization of matter multiplets. The remaining functional
dependence $X(S)$ will be fixed by loop-level consistency of theory.

\subsection{Orbifolding}

The analysis above has been performed for the case that corresponds
to the circle compactification of the fifth dimension. It is
straightforward to extend it to a compactification on the line
segment. For vector multiplets, what we have to do is removing a
bi-fundamental field, e.g.\ $Q_N$. This procedure leads to only a
four-dimensional vector multiplet as the light degrees of freedom. The
gauge anomaly arising from removing $Q_N$ can be supplemented by
introducing appropriate chiral fields which are charged under 
the $G_1$ and $G_N$ gauge symmetries. Effects of these `local' fields
can be neglected in the large $N$ limit. For matter multiplets, the
absence of one anti-chiral multiplet, e.g.\ $\bar\Phi_N$, leaves a
chiral zero mode of the fundamental representation of the diagonal
subgroup $G$. If one chiral multiplet, e.g.\ $\Phi_1$, is removed, a
zero mode is anti-fundamental representation. In these cases, suitable
anomaly cancellations are also required. The gauge-invariant
Lagrangian and its moduli dependences are almost same as before. For
example, when $\bar\Phi_N$ is removed, the mass eigenvalues and
eigenstates are given by
\begin{equation}
  m_n \;=\; m_{\psi_n} \;=\; 2v\sin\bigg(\frac{n\pi}{2N}\bigg) \qquad 
  (n=0,\cdots,N-1),
\end{equation}
\begin{equation}
  \widetilde\Phi_n = \sqrt{\frac{2}{2^{\delta_{n0}}N}}
  \sum_{j=1}^N\cos\left(\frac{2j-1}{2N}n\pi\right)\Phi_j, \qquad
  \widetilde{\bar \Phi}_n = \sqrt{\frac{2}{N}}
  \sum_{j=1}^{N-1}\sin\left(\frac{j}{N}n\pi\right)\bar\Phi_j.
\end{equation}
The result can be obtained for the case of removing $\Phi_1$ by
exchanging $\Phi\leftrightarrow\bar\Phi$, and in a similar way for
vector multiplets. The SUSY-breaking mass formulas do not change
except for the expression of $m_n$. It is, however, noted as a
difference from the previous section that bare mass parameters of
matter multiplets cannot be introduced unless there is a set of matter
multiplets which leaves vector-like massless modes.

\section{Quantum analysis}

We have discussed SUSY-breaking effects through the moduli $S$
and $Q$, and calculated the tree-level spectrum by determining the
proper moduli dependences of the action. We have particularly shown
that on the specific lines in the parameter space of $F_S$ 
and $F_Q$, the sparticle spectra are consistent to the SUSY-breaking
patterns in higher dimensions. In some cases, low-energy degrees of
freedom do not receive SUSY-breaking effects and remain massless in
the tree-level approximation. It is then important, e.g.\ for
realistic model building, to include radiative corrections. In this
section, we will calculate various types of one-loop corrections to
gaugino and Higgs scalar masses from gauge and Yukawa interactions. In
the following, the analysis will be performed in the case of the
orbifold compactification.
 
\subsection{Gaugino masses}

\subsubsection{Vector contribution}

The first type of one-loop corrections which contribute to zero-mode
gaugino masses involves vector and adjoint scalar multiplets running
in the loops. Let us recall that the mass matrix of the $n$-th massive
gauge fermions takes the form:
\begin{equation}
  -\frac{1}{2} \pmatrix{ \tilde\lambda_n & \tilde\chi_n }
  \pmatrix{ \frac{F_S}{2\langle S\rangle} & m_n \cr m_n & 
  -\frac{F_Q}{\langle Q\rangle} -\frac{F_S}{2\langle S\rangle}}
  \pmatrix{ \tilde\lambda_n \cr \tilde\chi_n } \,+{\rm h.c.},
\end{equation}
where $\chi_n$ are the goldstone fermions associated with the gauge
symmetry breaking. The eigenvalues $m_{\lambda_{n\pm}}$ of this matrix 
were given by (\ref{mlam}) and the mixing angle between $\lambda_n$
and $\chi_n$ upon the diagonalization satisfies
\begin{equation}
  \tan 2\theta_n \;=\; \frac{2m_n\langle S_5\rangle}{F_{S5}}. 
\end{equation}
At the component level, there are two types of one-loop diagrams for
the gaugino two-point function, which come from gauge fields and
adjoint scalars, respectively,
\begin{eqnarray}
  I_{\rm gauge} &=& \bigg(\frac{g}{\sqrt{N}}\bigg)^2C_2(G) 
  \sum_{n=0}^{N-1} \int\!\frac{d^4k}{i(2\pi)^4}\, 
  \frac{g_{\mu\nu}}{(p-k)^2-m_n^2}\, \sigma^\mu
  \Bigg[\frac{\cos^2\theta_n}{m_{\lambda_{n+}}\!\!-k\!\!\!/}
  +\frac{\sin^2\theta_n}{m_{\lambda_{n-}}\!\!-k\!\!\!/}\Bigg]
  \sigma^\nu, ~~~ \\[1mm]
  I_{\rm adj} &=& \bigg(\frac{\sqrt{2}g}{\sqrt{N}}\bigg)^2C_2(G)
  \sum_{n=0}^{N-1} \int\!\frac{d^4k}{i(2\pi)^4}\, 
  \frac{1}{m_{c_n}^2\!-(p-k)^2}
  \Bigg[\frac{\sin^2\theta_n}{m_{\lambda_{n+}}\!\!-k\!\!\!/}
  +\frac{\cos^2\theta_n}{m_{\lambda_{n-}}\!\!-k\!\!\!/}\Bigg],
\end{eqnarray}
where $p$ is the four-momentum of the external gaugino line 
and $C_2(G)$ is the quadratic Casimir for the adjoint representation
of the gauge group $G$. With the introduction of a ultraviolet 
cutoff $\Lambda$, the divergence parts are calculated as 
\begin{eqnarray}
  I_{\rm gauge}^{\rm div} &=& \frac{C_2(G)}{16\pi^2}
  \bigg(\frac{g}{\sqrt{N}}\bigg)^2 \sum_{n=0}^{N-1}
  \Big(p\!\!\!/ -4\big(m_{\lambda_{n+}}\cos^2\theta_n
  +m_{\lambda_{n-}}\sin^2\theta_n \big)\Big)\ln\Lambda^2, \\
  I_{\rm adj}^{\rm div} &=& \frac{C_2(G)}{16\pi^2}
  \bigg(\frac{g}{\sqrt{N}}\bigg)^2 \sum_{n=0}^{N-1}
  \Big(p\!\!\!/ +2\big(m_{\lambda_{n+}}\sin^2\theta_n
  +m_{\lambda_{n-}}\cos^2\theta_n \big)\Big)\ln\Lambda^2. 
\end{eqnarray}
These divergences should be renormalized by appropriate counter
terms. Note in particular that the total one-loop divergent correction
to gaugino mass becomes
\begin{equation}
  \delta m^{(1)} \;=\; \frac{2NC_2(G)}{16\pi^2}
  \bigg(\frac{g}{\sqrt{N}}\bigg)^2 \frac{F_{S5}}{\langle S_5\rangle}
  \ln\Lambda^2.
  \label{gaufin}
\end{equation}
It is interesting that since this total mass divergence is
proportional to $F_{S5}$, there appears no ultraviolet divergence in 
the limit of the radius modulus SUSY breaking. This is indeed a
consistent result and will be discussed in detail later on. On the
other hand, the divergences generally appear in other cases, for
example, in the dilaton/moduli dominant cases.

The finite part of the gauge field correction $I_{\rm gauge}$ is given by
\begin{equation}
  I_{\rm gauge}^{\rm fin} \;=\; \frac{-2NC_2(G)}{16\pi^2}
  \bigg(\frac{g}{\sqrt{N}}\bigg)^2 \frac{F_T}{\langle T\rangle}.
\end{equation}
We have used an approximation that the zero-mode gaugino mass, 
i.e.\ the SUSY-breaking mass scale is much smaller than KK masses. The
finite correction from $I_{\rm adj}$ can also be estimated in a
similar way and is found to be $-\frac{1}{2}I_{\rm gauge}^{\rm fin}$.
We thus obtain the zero-mode gaugino mass up to one-loop level, 
\begin{equation}
  m_{\lambda_{0\pm}} \;=\; \pm\frac{F_{S5}}{2\langle S_5\rangle}
  -\frac{F_T}{2\langle T\rangle}
  -\frac{2NC_2(G)g_4^2}{16\pi^2}\frac{F_T}{\langle T\rangle}.
  \label{mvec}
\end{equation}
The four-dimensional effective gauge coupling $g_4$ is defined in 
(\ref{match}). One can find two important points from this expression
of one-loop gaugino masses. Firstly, the radiative correction is
proportional to $F_T$ but not $F_S$. This fact seems to agree with a
result of the string-inspired supergravity models~\cite{sugra}. There
one-loop corrections to gaugino masses arise from string threshold
corrections or effects of bulk fields, and are specified by the
modulus field $T$. The second point is that the size of one-loop
correction is controlled by the gauge beta function of heavy
fields. This fact coincides with the spectrum of Ref.~\cite{KK} which
is determined by wavefunction renormalizations from vector
messengers~\cite{GR}. The dependence on beta-function coefficients
also appears in the supergravity models.

\subsubsection{Matter contribution}

Consider one-loop contribution to gaugino masses from matter
multiplets. With the mass splitting between matter fermions and
sfermions, the one-loop correction to the gaugino two-point function
is evaluated as
\begin{eqnarray}
  I_{\rm mat} &=& \bigg(\frac{\sqrt{2}g}{\sqrt{N}}\bigg)^2T_2(R) 
  \sum_{n=0}^{N-1} \int\!\frac{d^4k}{i(2\pi)^4} \Bigg[
  \frac{1}{m_n-k\!\!\!/}\bigg(
  \frac{1/2}{m_{\phi_{n+}}^2\!\!-(p-k)^2}+
  \frac{1/2}{m_{\phi_{n-}}^2\!\!-(p-k)^2}\bigg) \nonumber \\[1mm]
  && \hspace*{3.9cm} -\frac{1}{m_n+k\!\!\!/}\bigg(
  \frac{1/2}{m_{\phi_{n+}}^2\!\!-(p-k)^2}+
  \frac{1/2}{m_{\phi_{n-}}^2\!\!-(p-k)^2}\bigg)\Bigg],
  \label{I}
\end{eqnarray}
where $T_2(R)$ is the quadratic Dynkin index for the 
representation $R$ of the unbroken gauge group $G$. The scalar mass
eigenvalues $m_{\phi_{n\pm}}$ are given in Eq.~(\ref{stree}) 
with $\Gamma=1/4$ (see the discussions in Sections 2.2.2 
and 2.2.3). The divergent part of (\ref{I}) is given by
\begin{equation}
  I_{\rm mat}^{\rm div} \;=\; \frac{2NT_2(R)}{16\pi^2}
  \biggl(\frac{g}{\sqrt{N}}\biggr)^2 p\!\!\!/\,\ln\Lambda^2.
\end{equation}
Note that this is a supersymmetric correction. There is no divergence
to gaugino masses from matter multiplets since only scalar components
receive SUSY-breaking masses. In other words, the loop integrals
converge if the scalar propagators are expanded with respect to 
SUSY-breaking VEVs. As for the finite part, we obtain
\begin{equation}
  I_{\rm mat}^{\rm fin} \;=\; \frac{NT_2(R)}{16\pi^2}
  \bigg(\frac{g}{\sqrt{N}}\biggr)^2 \frac{F_T}{\langle T\rangle}.
  \label{m2}
\end{equation}
One can see that as in the vector case, the matter contribution is
controlled by the modulus $T$. After all, the total amount of mass
corrections is read from (\ref{mvec}) and (\ref{m2}), and the gaugino
mass up to one-loop level is given by
\begin{equation}
  m_{\lambda_{0\pm}} \;=\; \pm\frac{F_{S5}}{2\langle S_5\rangle}
  -\frac{F_T}{2\langle T\rangle} + N\Big[-2C_2(G)+2T_2(R)\Big]
  \frac{g_4^2}{16\pi^2}\frac{F_T}{\langle T\rangle}.
\end{equation}
The radiative correction is proportional to $F_T$ and the gauge beta
function of heavy supermultiplets.

\subsection{Higgs masses}

Next we study one-loop radiative corrections to masses of Higgs
scalars. Having the large top Yukawa coupling in mind, the corrections
from various types of large Yukawa couplings will be investigate in
detail. Gauge corrections can be estimated in a similar way and give
qualitatively similar results. In the numerical evaluations, we will
focus on the three special limits discussed in Section 2, and find
that all types of radiative corrections to Higgs masses are consistent
with the expected SUSY-breaking patterns. In particular, the
corrections become ultraviolet finite in the limit of the radius
modulus SUSY breaking. (We have already found in the previous section
the same result for gaugino masses [see (\ref{gaufin})].) \ The
compactification-scale dependence of radiative corrections is another
important factor to be examined. We will show that its quantitative
behavior is also properly taken into account in our framework. 

\subsubsection{Bulk-Brane-Brane couplings}

Let us study the radiative corrections from Yukawa couplings
involving KK modes of matter fields. The first case we consider is
a Yukawa coupling among one `bulk' field and two `brane'
fields. Suppose that $L_i$ ($i=1,\cdots,N$) 
and $\bar L_j$ ($j=1,\cdots,N-1$) are introduced as bulk matter
multiplets as described in Section 2.2 and $e$ as a chiral superfield
charged only under $G_1$ (would-be a localized field on a brane in the
five-dimensional point of view). The transformation properties are
listed below;
\begin{equation}
\begin{array}{c|ccccc}
  & ~G_1~ & ~G_2~ & ~G_3~ & ~\cdots~ & ~G_N~ \\ \hline
 L_1 & \f & 1 & \cdots & \cdots & 1 \\
 \bar L_1 & \af & 1 & \cdots & \cdots & 1 \\
 L_2 & 1 & \f & \cdots & \cdots & 1 \\
 \bar L_2 & 1 & \af & \cdots & \cdots & 1 \\
 \vdots & \vdots & \vdots & \vdots & \ddots & \vdots \\
 e & \af & 1 & \cdots & \cdots & 1 \\ 
 H & 1 & 1 & \cdots & \cdots & 1
\end{array}
\end{equation}
In addition to the tree-level matter Lagrangian (\ref{Lmat}), we have
a gauge-invariant Yukawa term among $L_1$, $e$ and the Higgs 
field $H$. Other gauge invariant couplings may be forbidden by
symmetry arguments, for example, an invariance under $e\to -e$ 
and $H\to -H$. The resultant superpotential is written in terms of the
mass eigenstates of $L$'s defined in Section 2;
\begin{equation}
  W \;=\; y\bigg(\sum_{n=0}^{N-1} \eta_n^L \widetilde L_n\bigg) eH 
  +\sum_{n=0}^{N-1} m_{L_n} \widetilde L_n \widetilde{\bar L}_n, 
  \label{W}
\end{equation}
where $m_{L_n}$ are the KK masses (\ref{spimass}), and $\eta_n^L$
denotes the mixing between $L_1$ and the mass 
eigenstate $\widetilde L_n$. Its explicit form has been derived in
Section 2.3:
\begin{equation}
  \eta_n^L \;=\; \sqrt{\frac{2}{2^{\delta_{n0}}N}}
  \cos\bigg(\frac{n\pi}{2N}\bigg).
\end{equation}
Eliminating the matter auxiliary fields from the Lagrangian
(\ref{Lmat}) and (\ref{W}) leads to
\begin{eqnarray}
  {\cal L} &=& -y\psi_e\bigg(\sum_{n=0}^{N-1}\eta_n^L 
  \psi_{\widetilde L_n}\bigg)H -\frac{|y|^2}{2} \bigg(\bigg|
  \sum_{n=0}^{N-1}\eta_n^L \widetilde{L}_n'\bigg|^2 +\bigg|
  \sum_{n=0}^{N-1}\eta_n^{L*} \widetilde{\bar L}_n'\bigg|^2\bigg) 
  H^\dagger H \nonumber \\ 
  && \qquad -\frac{y}{\sqrt{2}}eH\sum_{n=0}^{N-1}\eta_n^L \Bigg[
  \bigg(\frac{F_S}{2\langle S\rangle}-m_{L_n}^*\bigg)\widetilde{L}_n'
  +\bigg(\frac{F_S}{2\langle S\rangle}+m_{L_n}^*\bigg)
  \widetilde{\bar L}_n'\!{}^\dagger\Bigg] +{\rm h.c.} + \cdots, ~
  \label{bbb}
\end{eqnarray}
where $\psi_X$ means the spinor component of a chiral 
multiplet $X$. The sfermion fields with primes are the mass
eigenstates which completely diagonalize the mass matrix involving
SUSY-breaking effects,
\begin{eqnarray}
  \widetilde{L}_n \;=\; \frac{1}{\sqrt{2}}\widetilde{L}_n'
  +\frac{1}{\sqrt{2}}\widetilde{\bar L}_n'\!{}^\dagger, \qquad
  \widetilde{\bar L}_n \;=\; \frac{-1}{\sqrt{2}}\widetilde{L}_n'
  +\frac{1}{\sqrt{2}}\widetilde{\bar L}_n'\!{}^\dagger.
  \label{compL}
\end{eqnarray}
It is important to notice that in deriving the Lagrangian (\ref{bbb}),
we have fixed the normalization function $K_h$ for matter
multiplets as $\big\langle \frac{\partial \ln K_h}{\partial S}  
\big\rangle = \frac{1}{2\langle S\rangle}$. As will be seen below,
this form of VEV is required for the model to be consistent
at a quantum level. Together with Eq.~(\ref{kh}), we obtain
\begin{equation}
  K_h(S,S^\dagger) \;=\; c(S+S^\dagger)
  \label{kh2}
\end{equation}
with an arbitrary constant $c$. 

\begin{figure}[htbp]
\begin{center}
\leavevmode
\epsfxsize=10cm \epsfbox{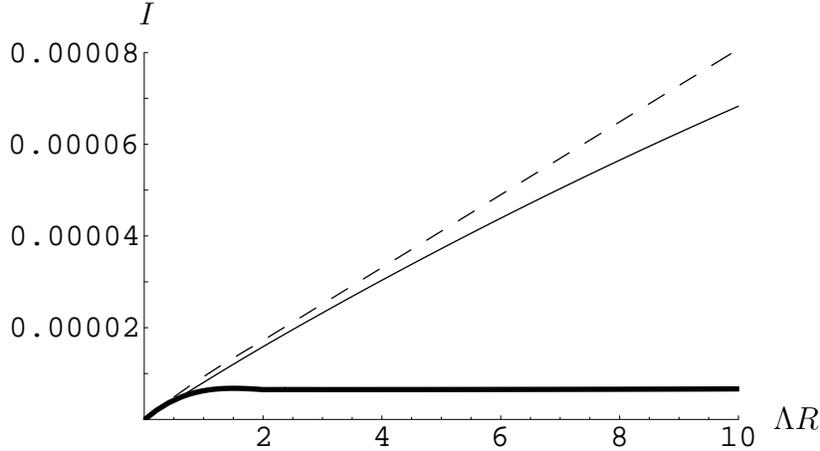}
\put(6,9){$\Lambda R$}
\put(-234,162){$I$}
\end{center}
\caption{The ultraviolet cutoff dependences of the loop correction
from the bulk-brane-brane type Yukawa coupling. From the upper, the
dilaton, moduli and the radius modulus SUSY breaking are plotted. We
take $y=1$ and $N=200$ as an example.} 
\label{bubrbr}
\end{figure}

With these matter-Higgs interactions at hand, we have three types of
diagrams which contribute to the Higgs mass corrections. Each diagram
involves an interaction in the Lagrangian (\ref{bbb}): (1) Yukawa
couplings with the fermions running in the loop, (2) sfermion-Higgs
quartic couplings, and (3) sfermion-Higgs trilinear couplings. These
contributions to the Higgs two-point function become
\begin{eqnarray}
  I^{(1)} &=& -2|y|^2 \sum_{n=0}^{N-1} \int\!\frac{d^4p}{i(2\pi)^4}
  \frac{|\eta_n^L|^2}{p^2-m_{L_n}^2}, \\[1mm]
  I^{(2)} &=& |y|^2\sum_{n=0}^{N-1}\int\!\frac{d^4p}{i(2\pi)^4}
  \frac{|\eta_n^L|^2}{2}\Bigg[\frac{1}{p^2-m^2_{L_{n+}}}
  +\frac{1}{p^2-m^2_{L_{n-}}}\Bigg] 
  +|y|^2\!\int\!\frac{d^4p}{i(2\pi)^4}\frac{1}{p^2}, \\[1mm]
  I^{(3)} &=& |y|^2 \sum_{n=0}^{N-1}
  \int\!\frac{d^4p}{i(2\pi)^4}\frac{|\eta_n^L|^2}{2}\left[
  \frac{\Big|\frac{F_S}{2\langle S\rangle}
  -m_{L_n}^*\Big|^2}{p^2-m^2_{L_{n+}}} +\frac{\Big|
  \frac{F_S}{2\langle S\rangle}+m_{L_n}^*\Big|^2}{p^2-m^2_{L_{n-}}}
  \right]\frac{1}{p^2}.
\end{eqnarray}
The sfermion mass eigenvalues $m^2_{L_{n\pm}}$ are given by
(\ref{stree}). We show in Fig.~\ref{bubrbr} the total one-loop
correction $I=I^{(1)}+I^{(2)}+I^{(3)}$. The horizontal axis denotes
the momentum cutoff $\Lambda$ normalized by the compactification
scale. We particularly focus on the three limits of the SUSY-breaking
order parameters which were presented in Table 1.
It is found from the figure that the loop correction is independent of
the momentum cutoff and becomes finite in the limit of the radius
modulus breaking (bold line). On the other hand, in the
dilaton/moduli dominance scenarios (dashed/solid lines), the
corrections linearly depend on $\Lambda$. This behavior is understood
as the number of KK modes running around the loop. That is, the KK
modes whose masses are below the momentum cutoff can contribute to the
radiative corrections. In these three limited cases, the structures
of radiative corrections are consistent with the results which have
been obtained in the literature.

The finite result for the Higgs mass is analytically understood
by noting that, in the limit $F_{S5}/\langle S_5\rangle
=F_S/\langle S\rangle+F_T/\langle T\rangle=0$, the total 
correction $I$ is simply summarized as
\begin{equation}
  I\,(\mbox{\small $F_{S_5}=0$}) \;=\; 
  -|y|^2 \sum_{n=0}^{N-1}\sum_{i=+,-}
  \int\!\frac{d^4p}{i(2\pi)^4} |\eta_n^L|^2\left[\frac{1}{p^2-m_{L_n}^2}
    -\frac{1}{p^2-m^2_{L_{n\,i}}}\right].
  \label{FS50}
\end{equation}
The first term in the integrand represents the fermionic contribution
and the second one the bosonic contribution. The ultraviolet
finiteness of this type of momentum integral will be discussed in
Section 5.

\subsubsection{Bulk-Bulk-Brane couplings}

\begin{figure}[htbp]
\begin{center}
\leavevmode
\epsfxsize=10cm \epsfbox{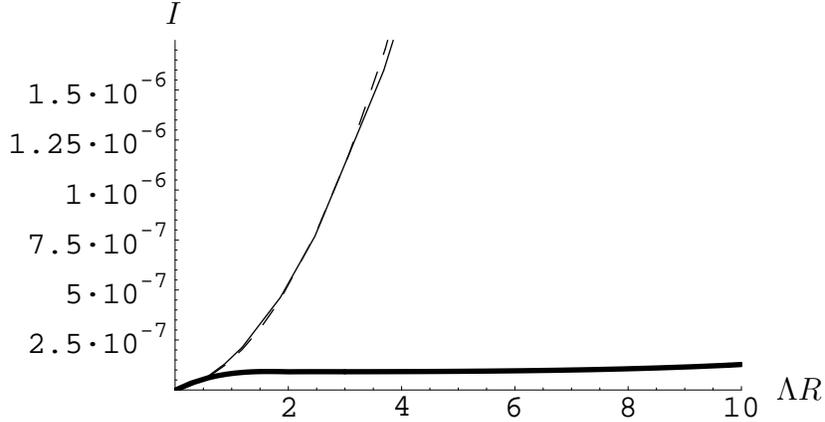}
\put(6,10){$\Lambda R$}
\put(-225,152){$I$}
\end{center}
\caption{The cutoff dependences of the loop correction from the 
bulk-bulk-brane type Yukawa coupling. The dilaton, moduli and radius 
modulus SUSY breaking are plotted from the above (the dashed, solid,
and bold lines, respectively). The parameters are same as in
Figure~\ref{bubrbr}.}
\label{bububr}
\end{figure}

In a similar way, we calculate one-loop Higgs masses from
bulk-bulk-brane type Yukawa couplings. In this case, the matter
multiplets $L_i$ ($\bar L_i$) and $e_j$ ($\bar e_j$) are charged under
the $G$ gauge symmetry, and the Higgs $H$ is introduced as a singlet
field. To implement the orbifold projection, $L_N$ and $\bar e_1$ are
removed from the spectrum. The transformation properties of these
fields are given by 
\begin{equation}
\begin{array}{c|ccccc}
  & ~G_1~ & ~G_2~ & ~G_3~ & ~\cdots~ & ~G_N~ \\ \hline
  L_1 & \f & 1 & \cdots & \cdots & 1 \\
  \bar L_1 & \af & 1 & \cdots & \cdots & 1 \\
  L_2 & 1 & \f & \cdots & \cdots & 1 \\
  \bar L_2 & 1 & \af & \cdots & \cdots & 1 \\
  \vdots & \vdots & \vdots & \vdots & \vdots & \vdots \\
  e_1 & \af & 1 & \cdots & \cdots & 1 \\ 
  e_2 & 1 & \af & \cdots & \cdots & 1 \\ 
  \bar e_2 & 1 & \f & \cdots & \cdots & 1 \\ 
  \vdots & \vdots & \vdots & \vdots & \vdots & \vdots \\
  H & 1 & 1 & \cdots & \cdots & 1
\end{array}
\end{equation}
The superpotential terms for masses and Yukawa couplings are written
in the mass eigenstate basis defined in a similar fashion to
(\ref{compL}),
\begin{eqnarray}
  W &=& y\bigg(\sum_{n=0}^{N-1} \eta_n^L \widetilde L_n\bigg)
  \bigg(\sum_{l=0}^{N-1} \eta_l^e \widetilde e_l\bigg)H 
  +\sum_{n=0}^{N-1} m_{L_n} \widetilde L_n \widetilde{\bar L}_n
  +\sum_{n=0}^{N-1} m_{e_n} \widetilde e_n \widetilde{\bar e}_n. 
\end{eqnarray}
Notice that we have introduced only one type of Yukawa 
term $yL_1e_1H$. This is analogous to the fact that a bulk Yukawa
coupling is not consistent with symmetries of five-dimensional
supersymmetric theories. In the present four-dimensional model, other
gauge-invariant Yukawa terms can be forbidden by some discrete
symmetry, as already done for the bulk-brane-brane type Yukawa
terms. Integrating out the matter multiplet $F$ terms, we obtain the
following interaction terms
\begin{eqnarray}
  {\cal L} &=& -\frac{|y|^2}{2}\Bigg[ 
  \bigg|\sum_{n=0}^{N-1}\eta_n^L \widetilde{L}_n'\biggr|^2 
  +\bigg|\sum_{n=0}^{N-1}\eta_n^{L*} \widetilde{\bar L}_n'\biggr|^2
  +\bigg|\sum_{n=0}^{N-1}\eta_n^e \widetilde{e}_n'\biggr|^2 
  +\bigg|\sum_{n=0}^{N-1}\eta_n^{e*} \widetilde{\bar e}_n'\biggr|^2
  \Bigg] H^\dagger H \nonumber \\
  && \quad -\frac{y}{2}H \sum_{n=0}^{N-1}\sum_{l=0}^{N-1} 
  \eta_n^e\eta_l^L (\widetilde{e}_n'-\widetilde{\bar e}_n'\!{}^\dagger) 
  \bigg[\Big(\frac{F_S}{2\langle S\rangle}-m_{L_l}^*\Big)
  \widetilde L_l' +\Big(\frac{F_S}{2\langle S\rangle} +m_{L_l}^*\Big)
  \widetilde{\bar L}_l'\!{}^\dagger\bigg] \nonumber \\
  && \quad -\frac{y}{2}H \sum_{n=0}^{N-1}\sum_{l=0}^{N-1} 
  \eta_n^L\eta_l^e (\widetilde{L}_n'-\widetilde{\bar L}_n'\!{}^\dagger)
  \bigg[\Big(\frac{F_S}{2\langle S\rangle}-m_{e_l}^*\Big)
  \widetilde e_l' +\Big(\frac{F_S}{2\langle S\rangle}+m_{e_l}^*\Big)
  \widetilde{\bar e}_l'\!{}^\dagger\bigg] \nonumber \\
  && \quad -y\bigg(\sum_{n=0}^{N-1}\eta_n^e \psi_{\widetilde e_n}\bigg)
  \bigg(\sum_{l=0}^{N-1}\eta_l^L \psi_{\widetilde L_l}\bigg)H 
  +{\rm h.c.} +\cdots. 
\end{eqnarray}
With this Lagrangian, the diagrams of loop corrections to the Higgs
mass are similar to those in the previous section: (1) Yukawa
couplings to the Higgs, (2) sfermion-Higgs quartic couplings, and (3)
sfermion-Higgs trilinear couplings. Each contribution is given by
\begin{eqnarray}
  I^{(1)} &=& -2|y|^2 \sum_{n=0}^{N-1}\sum_{l=0}^{N-1}
  \int\!\frac{d^4p}{i(2\pi)^4} |\eta_n^L|^2|\eta_l^e|^2
  \frac{p^2}{(p\!\!\!/-m_{L_n})(p\!\!\!/-m_{e_l})}, \\
  I^{(2)} &=& |y|^2\sum_{n=0}^{N-1}\int\!\frac{d^4p}{i(2\pi)^4}
  \frac{|\eta_n^L|^2}{2} \Bigg[\frac{1}{p^2-m_{L_{n+}}^2}
  +\frac{1}{p^2-m_{L_{n-}}^2}\Bigg] \;+ (L\leftrightarrow e), \\
  I^{(3)} &=& |y|^2 \sum_{n=0}^{N-1}\sum_{l=0}^{N-1}
  \int\!\frac{d^4p}{i(2\pi)^4}\frac{|\eta_n^L|^2}{2}\frac{|\eta_l^e|^2}{2}
  \Bigg[\frac{1}{p^2-m_{e_{n+}}^2}+\frac{1}{p^2-m_{e_{n-}}^2}\Bigg] 
  \nonumber \\
  && \hspace{35mm} \times \Bigg[\frac{\Big|\frac{F_S}{2\langle S\rangle}
  -m_{L_n}^*\Big|^2}{p^2-m_{L_{n+}}^2} 
  +\frac{\Big|\frac{F_S}{2\langle S\rangle}
  +m_{L_n}^*\Bigr|^2}{p^2-m_{L_{n-}}^2} \Bigg]
  \;+ (L\leftrightarrow e). 
\end{eqnarray}
The total loop correction $I=I^{(1)}+I^{(2)}+I^{(3)}$ is plotted for
several cases in Fig.~\ref{bububr}, where the Yukawa coupling
constant $y$ is taken to be $1$, as an example.
One can see that the cutoff dependences of the loop correction have 
similar behaviors as in the previous section; $I$ is insensitive to
the momentum cutoff $\Lambda$ for the radius modulus SUSY breaking,
while $I$ is cutoff-dependent (roughly proportional to $\Lambda^2$)
for the other cases, which dependence is interpreted as the number of
KK modes circulating in the loops.

\subsubsection{Bulk-Bulk-Bulk couplings}

Finally we consider the case that all $L$, $e$ and $H$ are introduced
as bulk multiplets, and examine radiative corrections to the massless
Higgs scalar. Suppose that corrections come from the superpotential
Yukawa coupling $L_1e_1H_1$ and other Yukawa terms are absent. Thus
the situation seems to be similar to the bulk-bulk-brane Yukawa case
in Section 3.2.2, except for an overall rescaling of Yukawa
couplings. However we now have an additional scalar 3-point vertex
which is a cross term generated by integrating out the $F$ component
of Higgs multiplet,
\begin{eqnarray}
  -y\sum_{n=0}^{N-1}\eta_n^H \bigg(\bigg\langle
  \frac{\partial \ln K_h}{\partial \ln S}\bigg\rangle
  \frac{F_S}{\langle S\rangle} \widetilde H_n+m_{H_n}^* 
  \widetilde{\bar H}_n\!\!{}^\dagger\bigg)
  \bigg(\sum_{n=0}^{N-1}\eta_n^L \widetilde L_n\bigg)
  \bigg(\sum_{l=0}^{N-1}\eta_l^e \widetilde e_l\bigg) ~+{\rm h.c.}.
  \label{new}
\end{eqnarray}
Other couplings of the Higgs zero mode are modified only by the
rescaling with the wavefunctions $\eta_0^{H,\,\bar H}=1/\sqrt{N}$,
which does not cause any qualitative changes to mass corrections.

While the corrections induced by the vertex (\ref{new}) are generally
divergent logarithmically, no qualitative change is found for the
supergravity SUSY-breaking models, compared to the
bulk-bulk-brane Yukawa case. On the other hand, one might wonder that
the vertex (\ref{new}) gives rise to a cutoff-dependent contribution
for the radius modulus breaking where even logarithmic divergences are
vanishing. However, it is found from the matter K\"ahler form
(\ref{kh2}) that the vertex (\ref{new}) leads to the interactions of
the Higgs $n=0$ modes
\begin{equation}
  -\frac{y}{\sqrt{2N}}\Big(m_{H_{0+}} \widetilde{H}_0'
  +m_{H_{0-}} \widetilde{\bar H}_0'\!{}^\dagger\Big)
  \bigg(\sum_{n=0}^{N-1}\eta_n^L \widetilde L_n\bigg)
  \bigg(\sum_{l=0}^{N-1}\eta_l^e \widetilde e_l\bigg) ~+{\rm h.c.},
\end{equation}
in the $F_{S5}=0$ limit. Therefore the coupling of the Higgs massless
mode of (\ref{new}) is vanishing, independently of which of the Higgs
fields $H$ and $\bar H$ contain the zero mode. As a result, the
radiative corrections from bulk-bulk-bulk Yukawa couplings are
qualitatively unchanged from the bulk-bulk-brane Yukawa case in the
previous section.

\subsection{Radius dependence}

We have analyzed the momentum cutoff dependence of the radiative
corrections. As another important factor, let us examine how the loop
corrections to scalar masses depend on the compactification radius $R$
of the fifth dimension. Consider for example the correction from a
bulk-brane-brane type Yukawa coupling. When all the parameters except
for $R$ are fixed, the following $R$ dependences of one-loop
corrections to scalar masses are expected;
\begin{eqnarray}
  \delta m^2_\phi &\propto& R \qquad [{\rm dilaton}], \label{expdi} \\
  \delta m^2_\phi &\propto& R^3 \quad\;\,\, [{\rm moduli}], \\
  \delta m^2_\phi &\propto& R^{-2} \quad\, [{\rm radion}]. \label{expra}
\end{eqnarray}
These behaviors can be understood as follows. In the dilaton dominant
case, the $R^1$ behavior is interpreted as the number of KK modes
($\simeq\Lambda R$) propagating in loop diagrams. In the moduli
dominant case, we have an $R^3$ factor from the number of KK modes 
($R^1$) and SUSY-breaking effects 
[$(F_T/\langle T\rangle)^2$ $\propto$ $R^2$]. On the other hand, the 
radiative corrections to Higgs masses are expected to behave like 
$R^{-2}$ in the radius modulus breaking. This is because the radius 
modulus breaking is equivalent to the boundary condition breaking of 
supersymmetry~\cite{GQ} and hence quadratic divergences are cut off 
by the compactification scale due to the locality in the extra 
dimension.

\begin{figure}[htbp]
\begin{center}
\leavevmode
\epsfxsize=10cm \epsfbox{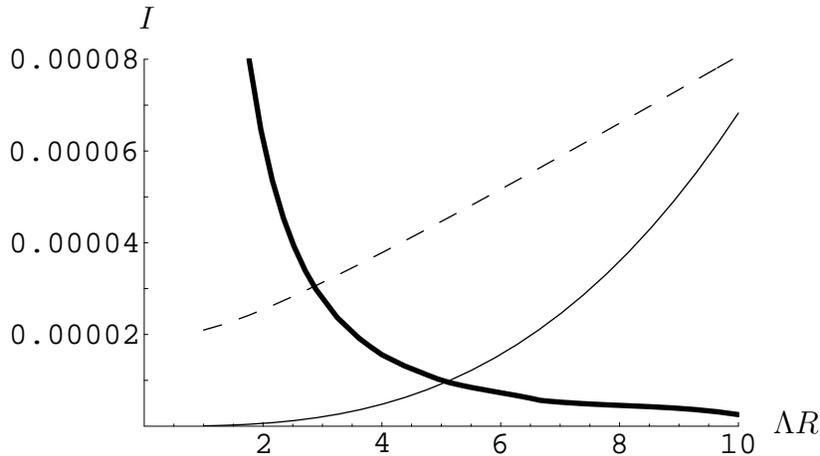}
\put(6,10){$\Lambda R$}
\put(-234,163){$I$}
\end{center}
\caption{The radius dependences of the radiative correction to the
Higgs mass. The horizontal axis means the compactification radius
(normalized by the ultraviolet cutoff) and the vertical one the
magnitudes of the correction. The dashed, solid, and bold lines
correspond to the dilaton dominance, the moduli dominance, and the
radius modulus breaking, respectively. The parameters are same as in
Figure~\ref{bubrbr}.}
\label{Rdep}
\end{figure}

In Fig.~\ref{Rdep}, we show the radius dependences of the scalar mass
correction from bulk-brane-brane Yukawa couplings (Section 3.2.1). The 
horizontal axis denotes the size of $R$ and the vertical one the 
magnitudes of one-loop corrections. In the figure, $R$ is taken to be 
within the regime $1<\Lambda R<10$ ($\Lambda$ is the momentum 
cutoff). The upper bound comes from the fact that finite scalar masses 
are realized in the region $\Lambda\ll v$ and the lower bound from 
the requirement that at least one KK mode runs around the loop. One 
can see from the figure that the expected results
(\ref{expdi})--(\ref{expra}) are certainly reproduced in our model. 

\bigskip

Summarizing Section 3, we have explicitly calculated one-loop 
radiative corrections to gaugino and Higgs scalar masses. The one-loop 
gaugino masses are proportional to the gauge beta function of heavy
fields and also to the modulus auxiliary VEV $F_T$. This seems to be a 
consistent result with the predictions of the string-inspired 
supergravity models. It is also important that the divergent
corrections to gaugino masses vanish in the limit of the radius
modulus SUSY breaking. For the Higgs mass, we have evaluated one-loop 
contributions from various types of Yukawa interactions. We have 
particularly found that in the limit of the radius modulus SUSY 
breaking, the corrections are insensitive to the ultraviolet cutoff of 
momentum integrals, while those related to the supergravity models 
depend linearly or quadratically on the cutoff scale. In addition, the
radius $R$ dependence of the Higgs mass has been studied. There we
have found that even for a finite number of gauge groups, that is,
with finite KK modes included, the consistent behaviors emerge in the
certain regions of modulus $F$ terms.

\section{Finiteness of radiative corrections}
\subsection{Large $N$}

We have observed by examining the cutoff and compactification scale 
dependences that finite radiative corrections appear in the limit of
the radius modulus breaking even for a finite number of gauge groups 
(i.e.\ finite KK modes). In continuous five-dimensional theory,
radiative corrections to Higgs masses are found to become finite if
one adopts SUSY breaking with boundary conditions~\cite{finite}. In
this case, it is important to include an infinitely large number of KK
modes in summing up loop corrections while preserving five-dimensional
supersymmetry. In other words, if one includes the effects of KK modes
which are sufficiently heavier than the momentum cutoff, the locality
in the extra dimension is recovered~\cite{KT}. In our model, that is
interpreted as involving a large number of gauge groups or taking a
lower four-dimensional cutoff. For example, for $N=200$, loop
corrections become cutoff independent in the regime $\Lambda R < 15$,
while the cutoff insensitivity cannot be seen for $N=10$. However we
here want to stress that to have finite corrections is not the main
point of this work because we only deal with a four-dimensional
theory. (For an approach to cutoff-insensitive Higgs masses in
product-group gauge theories, see~\cite{FGP}.)

\begin{figure}[htbp]
\begin{center}
\leavevmode
\epsfxsize=10cm \epsfbox{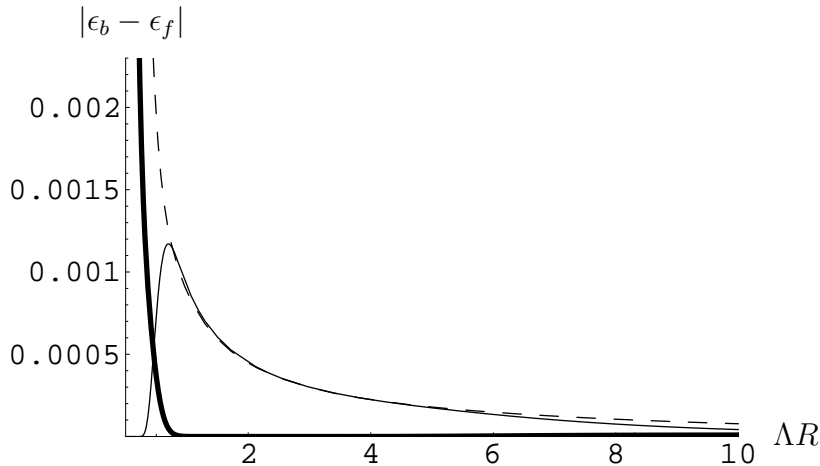}
\put(6,8){$\Lambda R$}
\put(-257,165){$|\epsilon_b-\epsilon_f|$}
\end{center}
\caption{The cutoff dependences of radiative corrections in the 
large $N$ limit. The horizontal axis denotes the cutoff scale and the
vertical one the difference between $\epsilon_b$ and $\epsilon_f$. We
take $F_S/\langle S\rangle$ and/or $F_T/\langle T\rangle$ $=1/10R$ in
the figure. The dashed, solid, and bold lines correspond to the
dilaton-, moduli-dominant, and the radius modulus SUSY breaking,
respectively.}
\label{delI}
\end{figure}

In this subsection, we analytically clarify the high-energy behavior 
of radiative corrections to Higgs scalar masses, especially focusing 
on the insensitivity to the momentum cutoff. Passing to Euclidean
momenta, consider the following typical form of radiative corrections
to Higgs masses from bosonic and fermionic KK modes:
\begin{eqnarray}
  I_b \;=\; |y|^2 \sum_{n,\,\pm} \int\!\frac{d^4p}{(2\pi)^4}
  \frac{1}{p^2+{m^2_n}_\pm}, \qquad I_f \;=\; 
  -2|y|^2 \sum_n \int\!\frac{d^4p}{(2\pi)^4} \frac{1}{p^2+m^2_n},
\end{eqnarray}
where the factor 2 in $I_f$ has been introduced so that the boson and
fermion degrees of freedom are equal. This type of corrections 
appears, for example, due to Yukawa couplings between bulk matter
multiplets and a boundary Higgs field. In fact, we have already
encountered it in Eq.~(\ref{FS50}). (That is a reason we have written
the coupling as $y$ in the above equations.) \ A similar analysis can
also be performed for gauge interactions. We treat these momentum
integrals with the proper-time regularization and then obtain for $I_b$
\begin{eqnarray}
  I_b \;=\; \frac{|y|^2}{16\pi^2}\sum_{n,\,\pm}\int^\infty_0\!
  dt \,\frac{1}{t^2}e^{-t{m^2_n}_\pm},
\end{eqnarray}
and a similar expression for $I_f$. An important point is that the
integration and summation can be safely exchanged if one introduces a
ultraviolet momentum cutoff $\Lambda$ and truncates the KK-mode sum at
a finite level. Note that the model in this paper naturally contains
a finite number of KK modes. Thus we find
\begin{eqnarray}
  I_b \;=\; \frac{|y|^2}{16\pi^2}\int^\infty_{1/\Lambda^2}\! dt\,
  \frac{1}{t^2} \sum_{n,\,\pm} e^{-t{m^2_n}_\pm}.
\end{eqnarray}
To see the cutoff dependence of radiative corrections, we examine the
beta functions in the sense of Wilsonian renormalization-group equations: 
\begin{eqnarray}
  \frac{\partial I_{b\,(f)}}{\partial \ln\Lambda^2} \;=\;
  \frac{\epsilon_{b\,(f)}\,|y|^2}{16\pi^2} \Lambda^2, 
  \hspace*{15mm} \\[3mm]
  \epsilon_b \;=\; \sum_{n,\,\pm} e^{-{m^2_n}_\pm/\Lambda^2}, \qquad
  \epsilon_{f} \;=\; 2\sum_n e^{-m^2_n/\Lambda^2}.
\end{eqnarray}
Since we are now interested in a particular relation between finite
radiative corrections and the radius modulus breaking, it is
appropriate to take a large $N$ limit (the five-dimensional limit) in
analyzing the corrections. In this limit, the results in Section 2
show that bosonic and fermionic KK modes have the following form of
tree-level mass eigenvalues
\begin{eqnarray}
  {m^2_n}_\pm \;=\; 
  \bigg(\frac{n}{R}\pm\frac{F_T}{2\langle T\rangle}\bigg)^2
  +\left|\frac{F_S}{2\langle S\rangle}\right|^2
  -\left|\frac{F_T}{2\langle T\rangle}\right|^2, \qquad
  m^2_n \;=\; \bigg(\frac{n}{R}\bigg)^2,
  \label{mass}
\end{eqnarray}
which is valid for vector and matter multiplets up to the second order 
of $F$'s (or exact in the dilaton/radius modulus SUSY-breaking 
cases). With this mass splitting, we show in Fig.~\ref{delI} the
differences between the beta function coefficients $\epsilon_b$ 
and $\epsilon_f$, i.e.\ the cutoff dependences of the radiative
corrections for the three typical cases.
It is clearly seen from the figure that the cutoff dependences are
rather different to each other, in particular, the ultraviolet
divergence is highly suppressed in the radius modulus breaking
case. The dilaton and moduli SUSY breaking have a similar behavior for
a large value of the cutoff scale $\Lambda$. This is because in these
cases the corrections are proportional to the number of KK modes, as
mentioned previously.

To analytically study these divergence properties, we estimate the 
contributions by use of the Poisson formula and obtain
\begin{eqnarray}
  \epsilon_b &=& 2\sqrt{\pi}\Lambda R
  \exp\Bigg[\frac{-1}{4\Lambda^2}\bigg(
  \bigg|\frac{F_S}{\langle S\rangle}\bigg|^2
  -\bigg|\frac{F_T}{\langle T\rangle}\bigg|^2\bigg)\Bigg]
  \sum_n e^{-\frac{\pi^2n^2}{(\Lambda R)^2}}\cos(\pi F_TR^2n), \\
  \epsilon_f &=& 2\sqrt{\pi}\Lambda R
  \sum_n e^{-\frac{n^2}{(\Lambda R)^2}}.
\end{eqnarray}
As can be seen in this expression, for a large value of $\Lambda R$,
the zero-mode contributions dominate due to the exponential dumping
factors. If assumed that the SUSY-breaking scale is much lower than
the compactification radius, the zero-mode contribution approximately
gives
\begin{eqnarray}
  \epsilon_b^0-\epsilon_f^0 &\simeq& \frac{\sqrt{\pi}}{2}\frac{R}{\Lambda}
  \Bigg(\left|\frac{F_T}{\langle T\rangle}\right|^2
  -\left|\frac{F_S}{\langle S\rangle}\right|^2\Bigg).
  \label{div}
\end{eqnarray}
The $\Lambda^{-1}$ dependence of (\ref{div}) implies
linearly-divergent corrections to Higgs masses. It can be seen from
(\ref{div}) that in the radius modulus breaking case 
where $F_{S5}\propto (F_S/\langle S\rangle+F_T/\langle T\rangle)=0$,
the radiative corrections become cutoff insensitive and therefore
finite results follow. On the other hand, the dilaton/moduli dominant
cases generally have divergent corrections proportional to $\Lambda R$.

\subsection{Boundary condition breaking}

In continuum five-dimensional theories, it is known that supersymmetry
breaking with boundary conditions gives finite radiative corrections
to scalar soft masses~\cite{finite}. This is partly because
supersymmetry is broken globally in the bulk and hence the breaking
effect is cut off by the size of the extra dimension. On the other
hand, in the present four-dimensional model, the finiteness is
achieved on a specific line of the SUSY-breaking parameter space. We
will show how these two cases are related. To this end, we formulate
the `boundary' condition breaking of supersymmetry in out setup and
examine whether the resulting spectrum matches to that of the radius
modulus breaking ($F_{S5}=0$). If these two spectra are found to be
equivalent, they may be transformed to each other by unitary
transformation, which gives a support that radiative corrections are
ultraviolet finite.

\subsubsection{Without twists}

Before discussing the boundary condition breaking, it is useful to
remind how to compactify the latticized extra dimensions. To begin
with, suppose that there are an infinite number of gauge groups and
corresponding matter fields. Compactifying physical space is performed
with the identification of indices under $i\to i+N$, which is
interpreted as a coordinate translation in the extra dimension. On the
other hand, the identification on the field space, i.e.\ boundary
conditions under this translation is given 
by $\Phi_{i+N}\equiv\Phi_i$. These procedures result in leaving 
the $N$ copies of gauge groups and matter fields as physical degrees
of freedom. The identification on the field space is clearly seen by
recombining the fields as
\begin{eqnarray}
  \Phi_j &=& \frac{1}{\sqrt{N}} \sum_{k=0}^{N-1} (\omega_k)^j\,
  \widetilde\Phi_k, 
\end{eqnarray}
where $\omega_k\equiv e^{\frac{2\pi k}{N}i}$, which is just the
Fourier mode expansion in the continuum limit. Note that there are
only finite numbers of mass eigenstates $\widetilde\Phi_k$, because
the model comes back to higher dimensions only in the large $N$
limit. At this stage, it is straightforward to include various types
of twisted boundary conditions. In what follows, we will study a
specific case which is related to the radius modulus breaking of
supersymmetry.

\subsubsection{$SU(2)$ twist}

Along the line discussed above, we demonstrate the twisted boundary
condition for vector multiplets as an illustrative example. The field
content we consider is given by Table (2.1) except that we now have an 
infinite numbers of gauge groups and $Q$ fields. In particular, there
are two types of spinors $\lambda$ and $\chi$ for every gauge
group. The identification of physical space is the same as the above
untwisted case. A new ingredient now introduced  is the boundary
condition on the field space. Let us consider the following $SU(2)$
twist upon the identification of field space;
\begin{equation}
  \pmatrix{ \lambda_{j+N} \cr \chi_{j+N} }
  \;=\; e^{2\pi i\, \theta_a\sigma_a} \pmatrix{ \lambda_j \cr \chi_j },
  \label{twist}
\end{equation}
where $\sigma_a$ are the Pauli's matrices. The parameters $\theta_a$
will turn out to be SUSY-breaking order parameters. The mass matrix of
these spinor fields must be consistent with this identification and
therefore takes the form (for small $\theta_a$)
\begin{equation}
  v\big( \lambda ~|~ \chi \big) (M_0+\delta M) 
  \left(\begin{array}{c}
    \lambda \\ \hline
    \chi
  \end{array}\right) +{\rm h.c.},
  \label{tm}
\end{equation}
\begin{eqnarray}
  M_0 &=& \left( \begin{array}{cccc|cccc} 
      & & & & 1 & & & -1 \\
      & & & & -1 & \ddots & & \\
      & & & & & \ddots & \ddots & \\ 
      & & & & & & -1 & 1\\ \hline
      1 & -1 & & & & & & \\ 
      & \ddots & \ddots & & & & & \\ 
      & & \ddots & -1 & & & & \\
      -1 & & & 1 & & & & 
    \end{array} 
  \right), \\
  \delta M &=& \left( \begin{array}{cccc|cccc} 
      & & &  & & & & -2\pi i\theta_3 \\
      & & & & & & & \\
      & & & & & & & \\ 
      & & & & & & & \\ \hline
      & & & & & & & -2\pi(i\theta_1+\theta_2) \\ 
      & & & & & & & \\ 
      & & & & & & & \\
      -2\pi i\theta_3 & ~~~~~ & & & -2\pi(i\theta_1+\theta_2) & & & 
    \end{array} 
  \right).
\end{eqnarray}
The $M_0$ part denotes the mass matrix for untwisted fields, whose 
eigenvalues are given by (\ref{KKmass}). On the other hand, $\delta M$ 
arises due to the twisting boundary condition (\ref{twist}). These
mass terms become the kinetic energy along the extra dimension if one
takes the continuum limit.

Let us diagonalize the mass matrix by perturbation theory with respect
to $\theta_a$. The eigenvalues and eigenmodes of the unperturbed 
matrix $M_0$ are given by
\begin{equation}
  m_{k\,\pm}^{(0)} \,=\, \pm 2v\sin\bigg(\frac{k\pi}{N}\bigg), \qquad\;
  \widetilde\psi_k^\pm \,=\, \frac{1}{\sqrt{2N}}\sum_{j=1}^N
  (\omega_k)^j(\pm\lambda_j+\chi_j).
\end{equation}
The irrelevant overall phases factors have been absorbed into field 
redefinition. In perturbation theory, the nonzero matrix elements 
of $\delta M$ linear in $\theta$'s are
\begin{equation}
  \big\langle \widetilde\psi_k^+\big|\,\delta M\,\big| 
  \widetilde\psi_k^+\big\rangle \;=\;
  \big\langle \widetilde\psi_k^-\big|\,\delta M\,\big| 
  \widetilde\psi_k^-\big\rangle \;=\;
  \frac{-2\pi}{N}\cos\bigg(\frac{2\pi k}{N}\bigg) 
  (i\theta_1 + \theta_2 + i\theta_3),
\end{equation}
and then the eigenvalues at the first order become
\begin{eqnarray}
  m_{k\,\pm}^{(1)} &=& m_{k\,\pm}^{(0)} -\frac{2\pi v}{N}
  \cos\bigg(\frac{2\pi k}{N}\bigg)(i\theta_1 + \theta_2 \pm i\theta_3). 
\end{eqnarray}
The first term on the right-handed side means the KK masses and
therefore the second one is interpreted as SUSY-breaking effect. For
example, in the case that $\theta_1, \theta_2\neq 0$ and $\theta_3=0$,
the mass eigenvalues for the low-lying modes ($k\ll N$) read
\begin{equation}
  m_{k\,\pm}^{(1)} \;\simeq\; \pm\frac{k}{R} -\frac{F_T}{2R},
  \label{su2ss}
\end{equation}
where the SUSY-breaking order parameter is identified 
as $F_T\equiv 2(i\theta_1+\theta_2)$. This is indeed the spectrum that
is given by the radius modulus breaking (\ref{gaurad}). Therefore the
finiteness property of radiative corrections in this case is now
understood.

For $\theta_3=0$, a half of supersymmetry in five dimensions is left
unbroken, namely, a linear combination of the zero-mode gauge fermions
is massless, which composes a four-dimensional massless vector
multiplet together with the gauge boson zero mode. When $\theta_3$ is
turned on as well as $\theta_{1,2}$, the remaining $N=1$ supersymmetry
is broken. This corresponds to applying the twisted boundary condition
with a $U(1)$ rotation which does not commute with the $N=1$
supersymmetry.

It may be interesting to rewrite the mass term (\ref{tm}) in terms of 
the fields with untwisted boundary conditions. We consider for
simplicity the twisting with $\sigma_2$. It is straightforward to
generalize the following result to the cases with generic types of
twisting. Let us redefine the spinor fields as
\begin{equation}
  \left(\begin{array}{c} 
    \lambda_j \\ \chi_j
  \end{array} \right) \;=\; e^{2 \pi i\theta_2 \frac{j}{N} \sigma_2} 
  \left(\begin{array}{c}
    \lambda'_j \\
    \chi'_j
  \end{array} \right).
\label{redef}
\end{equation}
It can be easily checked that the fields with primes satisfy the
untwisted boundary condition under the translation $j\to j+N$,
\begin{equation}
  \left(\begin{array}{c}
    \lambda'_{j+N} \\
    \chi'_{j+N}
  \end{array} \right) \;=\; 
  \left(\begin{array}{c}
    \lambda'_j \\
    \chi'_j
  \end{array} \right). 
\end{equation}
Almost all terms in the Lagrangian are invariant under this field
redefinition, while the mass term (\ref{tm}) is not. This means the
kinetic term along the fifth dimension is not invariant under the
coordinate-dependent phase rotation. In the basis of $\lambda'$ 
and $\chi'$, the mass matrix becomes for small $\theta$,
\begin{eqnarray}
  (\ref{tm}) &\simeq& \Big(\lambda' ~|~ \chi'\Big) M_0 
  \left(\begin{array}{c}
    \lambda' \\ \hline
    \chi'
  \end{array}\right)
  -\frac{2\pi \theta_2 v}{N} \pmatrix{ \lambda'_1 & \cdots & \lambda'_N} 
  \pmatrix{ & 1 & & \cr & & \ddots & \cr & & & 1 \cr 1 & & & } 
  \pmatrix{ \lambda'_1 \cr \vdots \cr \vdots \cr \lambda'_N} 
  \nonumber \\
  && \hspace{10mm}
  -\frac{2\pi \theta_2 v}{N} \pmatrix{ \chi'_1 & \cdots & \chi'_N } 
  \pmatrix{ & 1 & & \cr & & \ddots & \cr & & & 1 \cr 1 & & & } 
  \pmatrix{ \chi'_1 \cr \vdots \cr \vdots \cr \chi'_N }. 
\end{eqnarray}
Diagonalizing this matrix should lead to the eigenvalues (\ref{su2ss}) 
with $\theta_1=0$. The first term in the right-handed side is the
kinetic energy term along the fifth dimension. The second and third
terms can be interpreted as the mass terms which come from a VEV of
extra component of gauge field~\cite{hoso}. That is, these three terms
together make up the covariant derivative of the spinor fields. It was
shown~\cite{GQ} that the radius modulus breaking of supersymmetry is
equivalent to the Wilson line breaking, proving that a VEV of the
extra component of graviphoton field generates SUSY breaking in the
five-dimensional off-shell supergravity (as seen in the second and
third terms in the above). The present analysis makes it clear from a
four-dimensional viewpoint that the relevant VEV is that of the
modulus $Q$ ($F_Q\propto F_T$).

\section{Summary}

In this paper, we have formulated four-dimensional SUSY breaking in
product-group gauge theories. The model contains several modulus
fields corresponding to the dilaton and the radion. The modulus fields
satisfy the specific relations suggested by some correspondences to
higher-dimensional physics. From these, the relations are extracted
for the $F$-component VEVs of modulus fields. The non-trivial moduli
dependences of the action have also been determined. We have shown
that at intermediate energy regime, the mass spectra of typical SUSY
breaking scenarios (the dilaton/moduli dominance and the radius
modulus breaking) appear in the corresponding limits on the space of
SUSY-breaking order parameters. We have calculated in detail the
gaugino and Higgs scalar masses up to one-loop level. Our results seem
to be consistent with various aspects of bulk SUSY breaking, e.g.\ the
string-inspired supergravity models.

The cutoff dependences of loop corrections have been investigated in
detail. We have calculated the gaugino and Higgs mass corrections from
various types of gauge and Yukawa couplings. In particular, we have
shown that the insensitivity to ultraviolet cutoff emerges for the
radius modulus breaking case. However, for other cases, the spectrum
depends linearly or quadratically on the momentum cutoff. This can be
understood as the number of KK modes running in the loop
diagrams. The compactification radius dependence of the one-loop mass
spectrum has also been studied.

The finiteness property of radiative corrections has been examined
from several different viewpoints. In particular, we have formulated
in our setup the boundary condition breaking of supersymmetry (or the
Hosotani mechanism) and shown that the spectrum indeed agrees with
that of the radius modulus breaking case. This result indicates that
the obtained finite corrections are due to the global breaking of
symmetries in the bulk.

While, in this work, we have focused on the special limits
corresponding to higher-dimensional SUSY breaking, it will be 
interesting to investigate other regions of modulus $F$ terms. Such a
generic pattern of $F$ terms might induce sparticle mass spectra not
yet explored in the literature. It is also possible to extend the
present analysis to include brane matter which is interpreted as
being charged under one of the gauge groups $G_i$'s. The SUSY-breaking
masses of this type of fields depend on whether they can couple to the
moduli $S$ and $Q$. However there seems to be, in general, few
principles to fix modulus couplings of brane fields, and the
couplings would also depend on more fundamental theories. The presence
of brane fields may be useful to generate Yukawa hierarchy of quarks
and leptons. Explicit four-dimensional model construction along this
line leads concrete predictions of (super) particle spectrum, compared
to those of other models. The model parameters might then be
constrained by clarifying spectrum and applying it to supersymmetric
standard models, etc. We leave such phenomenological analyses to
future work.

\subsection*{Acknowledgments}

The authors would like to thank Tatsuo Kobayashi for a collaboration
at an early stage of this work and lots of valuable discussions. This
work is supported in part by the Special Postdoctoral Researchers
Program at RIKEN.


\end{document}